# Real-Space Magnetic Imaging of the Multiferroic Spinels $MnV_2O_4$ and $Mn_3O_4$

B. Wolin,[1] X. Wang,[1] T. Naibert,[1] S. L. Gleason,[1] G. J. MacDougall,[1] H. D. Zhou,[2,3]
S. L. Cooper,[1] and R. Budakian[1]

[1]*Department of Physics and Frederick Seitz Materials Research Laboratory,
University of Illinois, Urbana, Illinois 61801, USA*

[2]*Department of Physics and Astronomy,
University of Tennessee, Knoxville, Tennessee 37996-1200, USA*

[3]*National High Magnetic Field Laboratory,
Florida State University, Tallahassee Florida 32310-3706, USA*

**Abstract**
Controlling multiferroic behavior in materials will enable the development of a wide variety of technological applications.  However, the exact mechanisms driving multiferroic behavior are not well understood in most materials.  Two such materials are the spinels $MnV_2O_4$ and $Mn_3O_4$, where mechanical strain is thought to play a role in determining magnetic behavior.  Bulk studies of $MnV_2O_4$ have yielded conflicting and inconclusive results, due in part to the presence of mesoscale magnetic inhomogeneity, which complicates the interpretation of bulk measurements.  To study the sub-micron-scale magnetic properties of Mn-based spinel materials, we performed magnetic force microscopy (MFM) on $MnV_2O_4$ samples subject to different levels of mechanical strain.  We also used a crystal grain mapping technique to perform spatially registered MFM on $Mn_3O_4$.  These local investigations revealed 100-nm-scale "stripe" modulations in the magnetic structure of both materials.  In $MnV_2O_4$, the magnetization of these stripes is estimated to be $M_z \sim 10^5$ A/m, which is on the order of the saturation magnetization reported previously.  Cooling in a strong magnetic field eliminated the stripe patterning only in the low-strain sample of $MnV_2O_4$.  The discovery of nanoscale magnetostructural inhomogeneity that is highly susceptible to magnetic field control in these materials necessitates both a revision of theoretical proposals and a reinterpretation of experimental data regarding the low-temperature phases and magnetic-field-tunable properties of these Mn-based spinels.

**Introduction**

The wide variety of interactions and degrees of freedom in condensed matter systems yield some of the most complex and challenging problems in physics. When different types of order compete, materials can exhibit rich phase diagrams with linked structural, magnetic, and orbital ordering transitions. Two phenomena of great interest can result from this competition: multiferroism—the coexistence and coupling of different types of ferroic order (ferromagnetism, ferroelectricity, and ferroelasticity)—and magnetoresponsive behavior, i.e., large susceptibilities of physical properties to external perturbations, such as applied magnetic fields and pressure. Magnetoresponsive and multiferroic materials show great promise for practical applications, ranging from high-frequency actuators to precision sensors [1].

Various mechanisms can cause a coupling between magnetic and other primary order parameters [2-4], including the development of non-collinear spin order that breaks inversion symmetry [2,3], and the formation of multiferroic domains [4,5] and domain walls [6,7]. One of the grand challenges in the study of multiferroic and other magnetoresponsive materials has been to identify the specific magnetostructural and magnetoelectric mechanisms responsible for the different magnetoresponsive phenomena observed in numerous complex magnetic materials, including $ACuO_3$ ($A$=Se,Te) [8], Mn-doped $BiFeO_3$ [9], $EuTiO_3$ [10], $Y_2Cu_2O_5$ [11], $YbMnO_3$ [12], and the spinels $CoCr_2O_4$ [5], $MnCr_2O_4$ [13], $MnV_2O_4$ [14,15], and $Mn_3O_4$ [16-19].

The magnetic spinel family of compounds (chemical formula $AB_2X_4$)—which consists of an $A$-site diamond sublattice and a geometrically frustrated $B$-site pyrochlore sublattice [20]—is a particularly promising class of materials for studying the microscopic origins of magnetoresponsive behavior in magnetic materials. Magnetic spinels exhibit a range of diverse phases and phenomena that can be sensitively tuned using a variety of methods, including $A$- and/or $B$-site substitution, applied pressure, and/or applied magnetic field [5,13-19,21]. Due to the strong sensitivity of their physical properties to pressure and magnetic field, the magnetic spinels have important potential applications in catalysis, electrochemistry, and magnetic shape memory [22-27]. More broadly, magnetic domain formation is known to play a key role in raising the susceptibilities of complex materials to external perturbations [6,7,28,29]. However, the potential role of this mesoscale inhomogeneity on the magnetoresponsive properties of spinels has not been well investigated, because most previous research on the spinels has been conducted using bulk probes focusing on atomic length-scales such as neutron scattering [30-32], SQUID magnetometry [33-35], x-ray diffraction [33,36,37], and Raman scattering [38,39].

In this report, we explore the role of 0.1-10 μm scale magnetic inhomogeneity on the magnetic properties of two specific spinels, $MnV_2O_4$ and $Mn_3O_4$, using magnetic force microscopy (MFM). By using a sub-micron size magnetic probe, MFM can measure magnetic properties that are averaged over just tens of unit cells. Consequently, MFM measurements can reveal small-scale (0.1-100μm) magnetic inhomogeneities that have been overlooked in bulk measurements. We select the Mn-based magnetic spinels, $MnV_2O_4$ and $Mn_3O_4$, for study, because both materials exhibit similar magnetostructural properties and transitions at cryogenic temperatures that depend sensitively on the $B$-site constituent, V or Mn. For example, $MnV_2O_4$ is a cubic paramagnet at room temperature, and undergoes a magnetic transition to a collinear ferrimagnetic (FEM) configuration below T=57K. A second transition to a Yafet-Kittel (YK) type FEM configuration accompanied by a cubic-to-tetragonal structural transition occurs at T=53K [30,33,36,40]. By contrast, the cubic-to-tetragonal structural transition in $Mn_3O_4$ occurs at a significantly higher temperature, T=1440K, and the low-temperature magnetostructural phase behavior is more complex: $Mn_3O_4$ is a tetragonal paramagnet at room temperature and develops a triangular FEM configuration near T=42K. Near T=39K, an incommensurate spin ordering develops before $Mn_3O_4$ finally transitions to a cell-doubled YK-FEM magnetic phase with an orthorhombic crystal structure near T=33K [30-32,41].

In this study, we collected MFM images across a wide range of temperatures and magnetic fields from two samples of $MnV_2O_4$ with different levels of induced mechanical strain. We also studied MFM images from a single sample of $Mn_3O_4$ with inherent strain produced during crystal growth. Among a diverse range of magnetic patterns, we observe 100-nm scale "stripe" modulations in the magnetic structure present in the lowest-field phases of both materials. These stripe modulations are further organized into 1-10 μm scale domains associated with the local crystal structure. In $Mn_3O_4$, an observed correlation between stripe width and encompassing tetragonal domain size evidences a connection between mechanical strain and the magnetic patterns. In $MnV_2O_4$, we observe 100 nm-scale stripe modulations consistent with recent zero-magnetic-field TEM measurements of thin-foil $MnV_2O_4$ [42], and we find different magnetic behaviors in the high- and low-strain $MnV_2O_4$ samples. We also present a quantitative estimate of the local magnetization associated with these stripe domains in $MnV_2O_4$. We observe that modest applied magnetic fields (<30 kG) cause dramatic changes to—and the ultimate elimination of—the stripe domain patterns in both $Mn_3O_4$ and low-strain $MnV_2O_4$, but not in high-strain $MnV_2O_4$. These findings are consistent with theoretical results showing that mesoscale magnetic inhomogeneity can significantly lower the energy barrier for strain- and field-dependent phase changes in complex materials [28,29], and suggests that magnetic domain formation plays an important role in the magnetoresponsive behavior of these spinel materials.

## **Methods**

Single crystals of $MnV_2O_4$ were grown at the National High Magnetic Field Laboratory in Tallahassee using a traveling-solvent-floating-zone technique. Mixtures of MnO and $V_2O_3$ were ground, pressed, and calcined to form the seed and feed rods. A greater than stoichiometric amount of $V_2O_3$ was used to compensate for evaporation during growth. Details of the growth and characterization are reported elsewhere [33]. Single crystals of $Mn_3O_4$ were grown at the University of Illinois using a floating-zone technique. Commercially available $Mn_3O_4$ powder was pressed and sintered to form the feed and seed rods. The structural and magnetic properties of the resulting crystals are also reported elsewhere [16,41]. For both materials, crystallographic orientations were determined via room-temperature x-ray diffraction.

After characterization, the crystal surface normal to the [001] (cubic) direction was polished to <50nm roughness, and sputter coated with a 5nm layer of Au-Pd to dissipate static charge. Two $MnV_2O_4$ samples were prepared from the same growth. The first sample was a half-boule semicylinder measuring approximately 5mm × 2.5mm × 0.5mm. Epoxy was applied to the entire back surface of this sample, which was then attached to a sapphire backing-plate. The total thermal contraction occurring between the epoxy curing temperature and the base temperature used in this study ($T$=4K) is ten times larger for the epoxy than for the $MnV_2O_4$, and therefore significant mechanical strain is induced in the sample below $T$=77K [42]. A similar order-of-magnitude difference in thermal expansion coefficients between $MnV_2O_4$ foil and the Mo mount resulted in an estimated 0.03% compressive strain in $MnV_2O_4$ at 87 K and a <0.1% compressive strain near the cubic-to-tetragonal transition at 52K in $MnV_2O_4$ [42]. While this estimated compressive strain is less than the ~0.15% lattice striction measured in $MnV_2O_4$ at the cubic-to-tetragonal transition [37], it is large enough to influence domain formation in $MnV_2O_4$ [42]. The second $MnV_2O_4$ sample was a full-boule cylinder having a 5mm diameter and a 2mm length, and was specifically prepared to minimize mechanical strain below $T$=77K. This sample was attached to a copper backing-plate using a single point of epoxy at one edge, allowing the sample to thermally contract without interference from either the epoxy or backing plate. The increased sample thickness and single epoxy point mounting both act to minimize mechanical strain at the sample surface. Thermal contact between the sample and backing plate was maintained through the epoxy and physically through the sample-plate interface. In addition, long soak times (~10 minutes) were used to ensure thermal equilibrium was achieved.

Single crystals of $Mn_3O_4$ were grown at the University of Illinois using a traveling-solvent-floating-zone technique. To prepare the $Mn_3O_4$ sample, the $Mn_3O_4$ rod was diced into a rectangular block measuring approximately 1mm × 2mm × 1mm. The sample was polished normal to the [110] (tetragonal) direction and sputter coated with 1nm Au-Pd to prevent charging. The $Mn_3O_4$ sample was lithographically patterned with an array of unique location markers to provide spatial location information. We performed cryogenic electron backscatter diffraction (EBSD) experiments to determine the tetragonal crystal grain structure for comparison to MFM measurements. Using the location markers, we were able to align the magnetic and crystallographic data images with approximately 50nm accuracy, allowing us to correlate observed magnetic phenomena with the local crystal domain structure.

We performed low-temperature, frequency-modulated MFM using a $^4$He bath cryostat that had a built-in superconducting magnet. Data was collected in the temperature range from $T$=4.5K to $T$=80K and the magnetic field range from $B$=0T to $B$=3T. In all cases, the magnetic field was oriented normal to the sample surface, resulting in $B$ parallel to [001] (cubic) for both $MnV_2O_4$ samples and $B$ parallel to [110] (tetragonal) for the $Mn_3O_4$ sample. Commercially available atomic force microscopy cantilevers were evaporatively coated with a 10-nm thick layer of FeCo to provide magnetic sensitivity. With probe-sample separations of approximately 100 nm and scan rates as low as 100 nm/s, we were able to achieve a spatial resolution of approximately 50 nm for magnetic features. The cantilevers used in these experiments have resonance frequencies approximately $f_0$~25kHz, spring constants approximately $k$~0.3N/m, and quality factors approximately $Q$~350,000 at $T$=4K in vacuum. We measured the cantilever displacement interferometrically using a 1510nm laser in a fiber-optic Fabry-Pérot configuration [43], and we measured the cantilever frequency using a phase-locked loop (see Supplementary Section [44]).

To extract quantitative information from the $MnV_2O_4$ image data, we conducted a calibration experiment to characterize the magnitude and orientation of the magnetic moment of the MFM probe. A 70-nm thick, 70-µm long straight rectangular gold wire was patterned onto a Si substrate using electron-beam lithography and thermal evaporation. The wire measured 4µm wide for half the length and 1µm wide for the other half, with a step-like junction at the center (Figure S2). We calculated the magnetic field produced by an electric current running through this simple geometry using a finite-element electromagnetic solver. For areas far from the junction, the simulation results showed near-perfect agreement with analytical calculations for an infinite wire. To ensure maximum remnant magnetization, the ambient magnetic field in the cryostat was cycled up to $B$=3T and back to $B$=0T before any measurements were performed. With a constant 5mA current running through the wire, we recorded MFM frequency shift data in the area near the junction. Comparing this data with the calculated field curvature, we extracted the point spread function (PSF) of the MFM probe. This function is independent of the sample being scanned, and can be used to quantitatively analyze the MnV$_2$O$_4$ data because it relates the measured MFM frequency shift directly to the magnetic field curvature produced by the sample [45]. See the Supplementary Section [44] for more details.

## Results

Figure 1(a) shows MFM data collected from a region of the high-strain MnV$_2$O$_4$ sample after cooling from $T=70$K to $T=40$K, well into the YK phase [33,35,50], in the presence of a weak magnetic field, $B=3$kG. The approximate cubic lattice directions (white arrows and text in Figure 1(a)) were determined using room-temperature x-ray diffraction. We observe a space-filling magnetic patterning with domain and subdomain structures. Large (µm-scale) domains of predominantly positive (blue) or negative (red) frequency shift contain and define the boundaries of 100-nm scale stripe modulations. The large domains correspond to areas of well-defined stripe direction. Additionally, the stripe pitch, amplitude, and offset vary continuously across domains, as seen in Figure 1(d), which shows frequency shift data along the indicated line-cut (yellow dashed line, Figure 1(a)). The pitch variation in Figure 1(d) is only approximately 14%, but the pitch variation between the left-most and right-most domains is as large as 60%. The stripe pitch is anti-correlated between domains: in the boxed region of Figure 1(a), the modulation pitch in the blue domain is highest and the modulation pitch in the two red domains is lowest, indicating a likely influence of mechanical strain on the magnetic patterning. By calculating the standard deviation ($\sigma$) of the frequency shift data from an entire MFM scan, we measure the degree of magnetic inhomogeneity. Figure 1(b) plots $\sigma$ versus temperature for data collected during a zero-field cool of the high-strain $MnV_2O_4$ sample. We observe a sharp onset of magnetic inhomogeneity near $T=58K$ and a peak at $T=54K$. The degree of inhomogeneity distinctly decreases between $T=54K$ and $T=49K$, and at $T=49K$ the MFM images show a clear change in the magnetic patterning. Both the raw MFM data and the derived $\sigma$ vs. $T$ data clearly indicate two magnetic phase transitions in $MnV_2O_4$, consistent with previous reports [30,33,36,40]. Furthermore, the results shown in Figure 1(b) are qualitatively similar to measurements of the bulk magnetization [30,33,36,40]. The correlation between bulk magnetic behavior and 0.1-10µm scale magnetic inhomogeneity suggests that the low-temperature magnetic behavior of $MnV_2O_4$ can be well characterized by magnetic domain formation and heterogeneity. The observed subdomain structure explains the sharp drop in overall inhomogeneity observed below T=54K. Without a subdomain structure, we would expect the magnetic inhomogeneity to increase monotonically with decreasing temperature. These conclusions will be further explored in the discussion section.

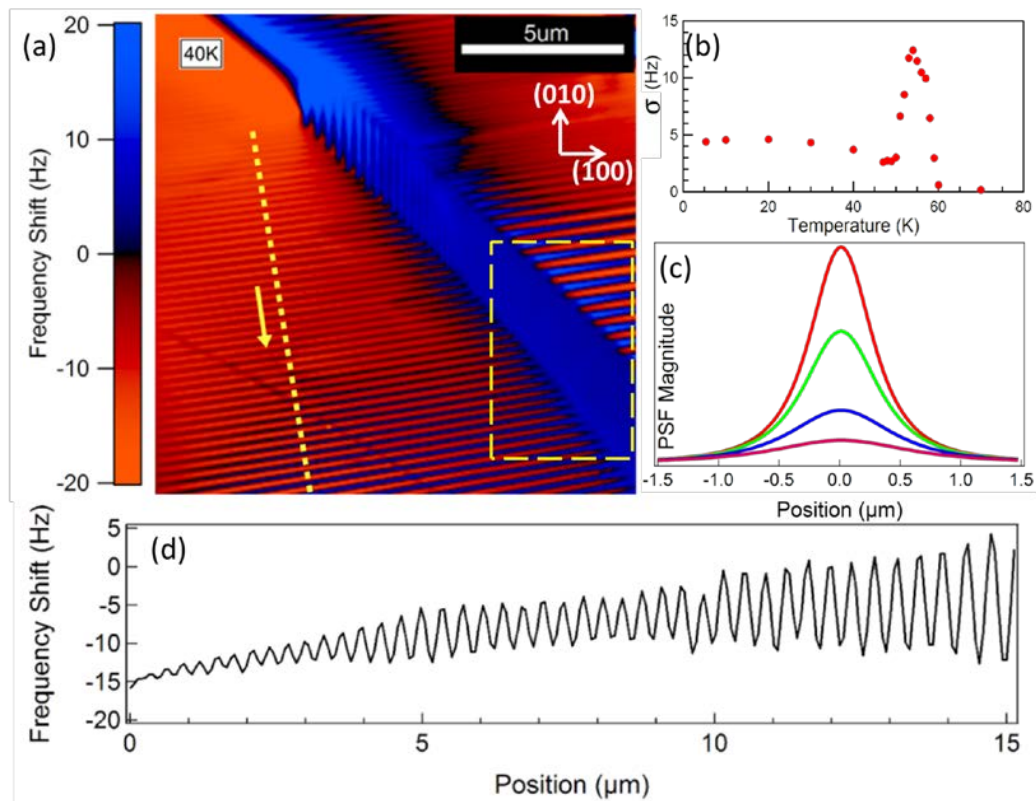

Figure 1: MFM data of high-strain MnV$_2$O$_4$ cooled from 70K to 40K in $B=0.3$T. (a) When cooled in a weak magnetic field, the magnetic pattern sharpens dramatically. We observe 20µm-scale domain structure with regular sub-domain stripe modulations. Regions of overall frequency shift (predominant blue or red color) correspond to areas of a single stripe direction. Approximate cubic lattice axes are indicated in white. The yellow dashed box highlights a region where mechanical strain influences the magnetic pattern. (b) Average magnitude of magnetic inhomogeneity (characterized by the standard deviation of frequency shift) measured while cooling the high-strain MnV$_2$O$_4$ sample in zero magnetic field. Note the qualitative similarity to measurements of the bulk sample magnetization. (c) Cross sections of the measured point spread function at locations: (top to bottom) 0nm, 250nm, 500nm, and 750nm away from the PSF center. (d) Frequency data along the indicated line through the 2-D image. Stripe pitch, frequency offset, and amplitude vary across the domain.

To make a quantitative comparison between the magnitude of magnetic inhomogeneity observed in MFM and the bulk magnetic behavior reported for $MnV_2O_4$, we performed a calibration experiment using previously established techniques [46-49]. Further details of the calibration experiment are included in the Supplemental Section [44]. Figure 1(c) shows the instrument response of the magnetic probe extracted from measurements of the calibration sample. From top to bottom, the traces show cross sections of the PSF at locations 0nm, 250nm, 500nm, and 750nm away from the probe apex. Using this measured spatial response function of the MFM probe, we quantitatively modeled the stripe pattern seen in Figure 1(a) to yield an estimate of the local magnetization associated with the sub-domain stripe features. We estimate (to within a factor of 3) the peak-to-peak magnetization associated with the stripe modulations to be $M_{pp} \approx 0.8 \cdot 10^5$ A/m. Because a cantilever-based magnetic probe is sensitive only to the magnetic field curvature, the absolute magnetization of a macroscopic sample cannot be determined using MFM; only gradients in the sample magnetization induce a frequency shift. Thus, our observations are consistent with two extreme possible interpretations: the stripes define regions with magnetization alternating either between $M_z=\pm M_{pp}/2$ or between $M_z=0$ and $M_z=M_{pp}$. Magnetometry experiments on $MnV_2O_4$ at $T=40K$ show that the bulk saturation magnetization is $M_z=0.7 \cdot 10^5$ A/m [35], so the magnetization associated with the stripe features is comparable to the overall magnetic behavior of the sample in both extreme cases. From these results, we conclude that the highly inhomogeneous nature of the magnetic state of $MnV_2O_4$ represents a dominant contribution to the magnetization that must be taken into account when analyzing the low-temperature magnetic behavior of this material.

Figure 2 shows representative MFM frequency shift data collected after cooling the low-strain MnV$_2$O$_4$ sample to $T$=40K in the presence of different magnetic field strengths. For fields in the range 0kG<$B$<2.5kG, we observe irregular magnetic patterning with large frequency shifts. Repeated cools with the same parameters yielded qualitatively distinct results, some with no regular patterning and others with highly regular stripe patterns. The observation that different cools yield different patterns indicates the existence of multiple, nearly degenerate metastable pattern states and the absence of significant pinning effects. Figure 2(a) shows an example of irregular patterning observed on cooling in zero applied field. In the field range 2.5kG<$B$<7.5kG, we observed 10µm-scale domain features oriented approximately 45° relative to the cubic crystal axes. We also observed sub-domain stripes that form an interwoven pattern, as can be seen in Figure 2(b). Repeated cools in this field regime with the same parameters yielded the same domain structure, but different sub-domain patterns. As the field is increased further, the number of sub-domain stripes decreases until only the domain features remain (Figure 2(c)). Between $B$=15kG and $B$=30kG (Figure 2(d)), all magnetic features are eliminated, indicating that the entire sample is a homogeneous magnetic domain.

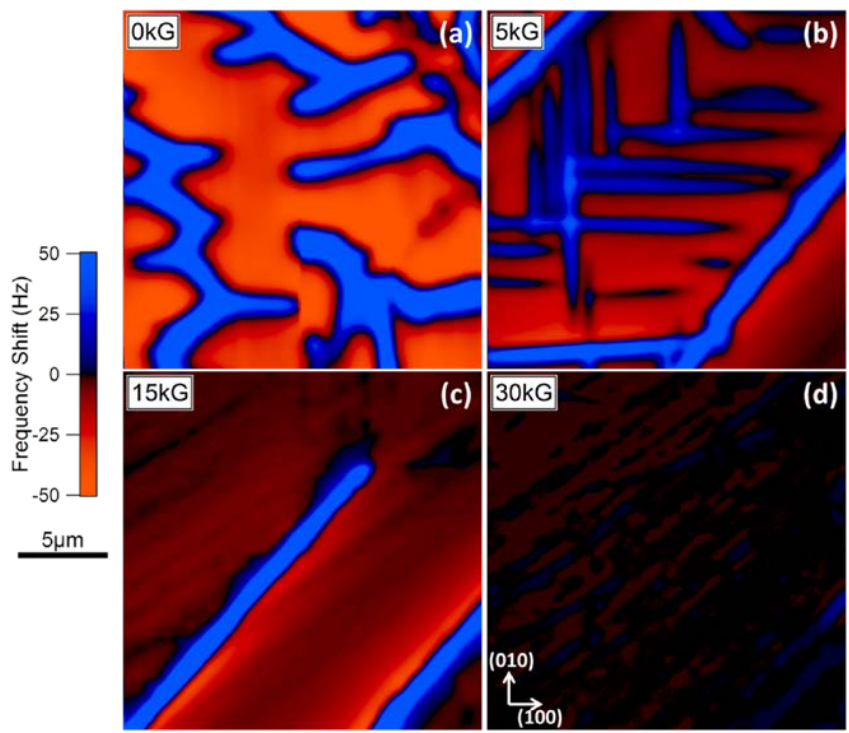

Figure 2: MFM data of low-strain MnV$_2$O$_4$ cooled to 40K. Images are 20x20µm. The approximate cubic axes in (d) apply to all panels. (a) At low fields, the magnetic pattern is amorphous and causes large frequency shifts. In addition, we observed several distinct types of patterning during different cools at the same field value. (b) In the intermediate field regime, 10µm-scale magnetic domains were observed. Irregular sub-domain striping was observed in tweed patterns. (c) At 10kG, sub-domain striping was eliminated. (d) By 30kG, all magnetic contrast was eliminated, indicating a single magnetic domain.

In the context of published phase diagrams [35,50], the temperature of the above measurements should place the material well within the tetragonal/YK phase for MnV$_2$O$_4$ for the entire field range investigated. The disappearance of magnetic features between B=1.5kG and B=30kG is consistent with reports of a weak first-order transition associated with the realignment of tetragonal domain structure [33,50], a conclusion supported by x-ray scattering measurements [36].

Consistent with this interpretation, we identify the strong domain features in Figures 2(b) and 2(c) as transitions between magnetic domains with magnetizations oriented parallel to different crystal axes (see Supplemental Section [44]), mirroring previously measured structural domains [36]. As the external magnetic field is increased, tetragonal domains not oriented parallel to the external field become energetically unfavorable, resulting in the magnetic uniformity shown in Figure 2(d).

Figure 3 shows representative MFM data collected while cooling the high-strain $MnV_2O_4$ sample to $T=40K$ in the presence of different magnetic field strengths. At low fields, we again observe irregular magnetic patterning, as shown in Figure 3(a). For fields $3kG<B<7.5kG$, we observe a less clearly delineated domain structure, as well as single direction sub-domain stripes, as shown in Figure 3(b). Finally, for $B>7.5kG$ (Figures 3(c,d)), a somewhat more complex magnetic patterning develops; this patterning changes as the magnetic field is increased, and includes the development of subdomain 100 nm-scale stripe features. Figure 3(d) shows that strong magnetic inhomogeneity persists up to the highest field measured, $B=30kG$. Though these measurements nominally explore the same region of phase space as those is Figure 2, the current results reveal a significant distinction between the high- and low-strain sample behaviors: high mechanical strain in the crystal lattice of $MnV_2O_4$ stabilizes magnetic inhomogeneity in higher magnetic fields. The distinct difference in magnetic domain patterns observed in the high-strain and low-strain samples also indicates a strong structural component to the magnetic domain pattern in $MnV_2O_4$. This connection could be further explored using a combination of MFM and local structural measurements, similar to that described below.

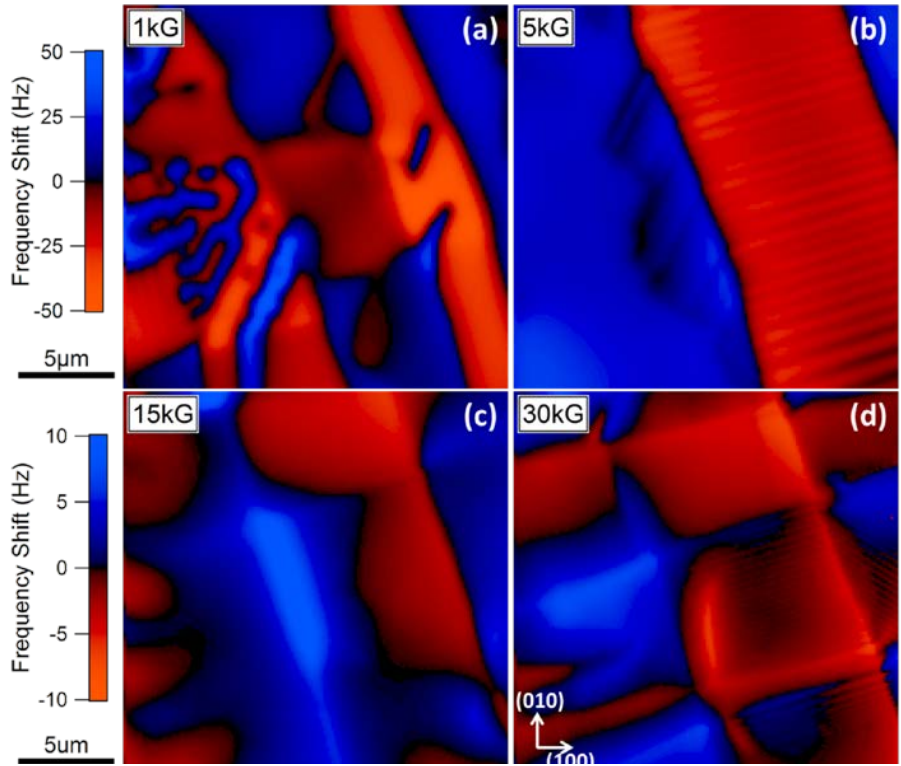

Figure 3: MFM data of high-strain $MnV_2O_4$ cooled to 40K. Images are 20x20µm. The approximate cubic axes in (d) apply to all panels. (a) At low fields, the magnetic pattern is also amorphous (similar to the low-strain measurements) and induces large frequency shifts. (b) In the intermediate field regime, a pattern of domains and sub-domain stripes appeared. No interwoven striping was observed at any field value. (c) At 15kG, the domain pattern becomes more segmented, but retains the features seen at lower fields. Stripe modulations are still present, but are difficult to observe due to large frequency shifts between domains. (d) Strong magnetic inhomogeneity remains at $B=30kG$, in contrast to the low-strain sample. The stripe modulations also persist up to $B=30kG$.

In an effort to investigate whether magnetic domain formation is observed in other Mn-based spinels exhibiting magnetoresponsive properties, we also used MFM to investigate the spatial organization of magnetic patterns in the magnetodielectric spinel, $Mn_3O_4$. Figure 4 is a composite MFM image of the $Mn_3O_4$ sample created by stitching together multiple individual MFM scans recorded in succession. The $Mn_3O_4$ sample was cooled in the presence of a weak magnetic field, $B$=2kG from above $T$=40K to $T$=18K; this is well into the cell-doubled orthorhombic ferrimagnetic phase, as determined by previous measurements [34, 41, 52]. We observe stripe modulations very similar to those observed in $MnV_2O_4$. In $Mn_3O_4$, the stripes form a tweed pattern consisting of different regions of coordinated stripe direction. The green dashed lines in Figure 4 indicate boundaries between the frozen-in tetragonal crystal grains, as determined by electron backscatter diffraction (EBSD). We observe a clear correspondence between the locations of tetragonal domain boundaries and the magnetic stripe region boundaries. Repeated cooling using the same parameters yields an identical set of magnetic domain boundaries, indicating that the magnetic domains are strongly pinned to the tetragonal crystal boundaries, similar to the behavior observed in the high-strain $MnV_2O_4$ sample. Furthermore, the size of the tetragonal domain is correlated with the stripe pitch within the domain in the $Mn_3O_4$ sample, with the largest tetragonal domains supporting stripes with the lowest pitch. As the tetragonal domain size shrinks, the stripe pitch increases until the MFM probe cannot resolve individual stripe features. Similar to our observations in $MnV_2O_4$, the tweed stripe pattern in $Mn_3O_4$ is eliminated by cooling in a sufficiently strong magnetic field ($B$=20kG). This is consistent with the observation of nearly degenerate orthorhombic phases in $Mn_3O_4$, and the selection of a universal orthorhombic distortion axis with applied field [34, 52]. The relationship between the tetragonal domains and the magnetic pattern is further evidence of the important role that mechanical strain plays in the low-temperature magnetic stripe formation and magnetic properties of these Mn-based spinels. The presence, magnitude, and similar field-behavior of magnetic inhomogeneities in both $Mn_3O_4$ and $MnV_2O_4$ indicate that such features are likely generic to a wider range of strongly spin-lattice coupled materials, particularly other magnetic spinels and magnetodielectric materials.

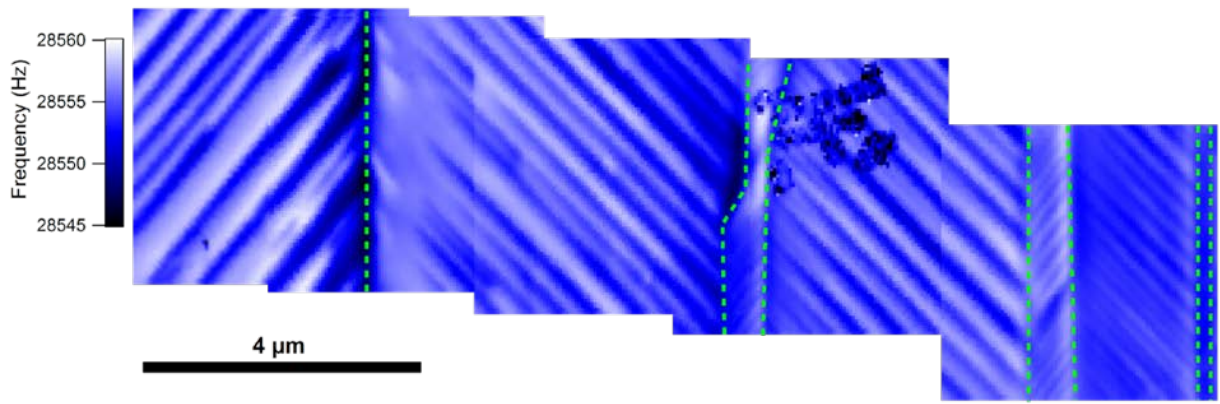

*Figure 4: Composite MFM image of $Mn_3O_4$ at T=18K, B=2kG. We observe tweed-pattern magnetic stripe features defined by the tetragonal crystal grain pattern (dashed green lines). The stripe widths are correlated to the domain size, suggesting a connection between the mechanical strain and the associated magnetic pattern. The patchy region in the second subpanel from the right reveals one of the location markers used to spatially register MFM data with EBSD results. The non-magnetic marker material does not affect the magnetic behavior of the sample, but appears in the data images because of the changing topography.*

## **Discussion**

Our investigations represent the first observations of nanoscale inhomogeneity in the low-temperature magnetic structures of bulk $MnV_2O_4$ and $Mn_3O_4$. Quantitative estimates of the magnetization associated with these nanoscale magnetic patterns indicate that the magnitude of the magnetic modulations is large, accounting for much of the bulk magnetic behavior reported in these materials. Additionally, our results show for the first time that the magnetic stripe modulations change significantly in modest magnetic field strengths that are comparable to the field strengths at which large magnetodielectric and magnetic-lattice striction effects are observed in $MnV_2O_4$ and $Mn_3O_4$ [30,36,37].

The nanoscale magnetic inhomogeneity we observe in $MnV_2O_4$ and $Mn_3O_4$ raises two fundamental questions: (i) what, if any, underlying structural inhomogeneity accompanies the magnetic inhomogeneity; and (ii) to what extent does the magnetic inhomogeneity contribute to the magnetoresponsive phenomena observed in $MnV_2O_4$ and $Mn_3O_4$ [16-18]?

Addressing the first issue, substantial direct and indirect evidence indicates that the nanoscale magnetic inhomogeneity we observe at low temperatures in $MnV_2O_4$ and $Mn_3O_4$ is associated with an underlying structural modulation. Bulk x-ray diffraction measurements on polycrystalline $Mn_3O_4$ [51] show evidence for a mixture of tetragonal and orthorhombic phases, and the coexistence of tetragonal (paramagnetic) and orthorhombic phases at low temperatures in $Mn_3O_4$ is also supported by recent muon spin resonance measurements of single-crystal $Mn_3O_4$, which reveal a mixture of magnetically ordered and disordered volumes at low temperatures [21]. The phonon and magnon Raman scattering spectra of heavily twinned samples of $Mn_3O_4$ also show evidence for phase coexistence at low temperatures, which may include coexisting orthorhombic and tetragonal phases [38]. More recent Raman experiments of the phonon and magnon spectra of untwinned $Mn_3O_4$ samples show clear evidence for coexisting face-centered orthorhombic and cell-doubled orthorhombic phases at low temperatures [52], consistent with the presence of a mesoscale structural modulation in this material. In $MnV_2O_4$, TEM measurements revealed the coexistence of tetragonal twinning domains with different c-axis orientations [42], and the sensitivity to strain we observe in our measurements of $MnV_2O_4$ support the conclusion that the nanoscale magnetic modulation we observe in this material is associated with an underlying structural modulation. Altogether, these results provide strong evidence that the magnetic modulations observed with MFM in both $MnV_2O_4$ and $Mn_3O_4$ are associated with an underlying structural modulation that betrays the strong coupling of spin, orbital, and structural degrees of freedom in these materials [36,37].

Notably, mesoscale magnetostructural modulations have been observed in other magnetic materials exhibiting strong spin-lattice coupling, including $La_{1.99}Sr_{0.01}CuO_4$ [53], $Co_{0.5}Ni_{0.205}Ga_{0.295}$ [54], and the Mn-doped spinel $CoFe_2O_4$ [55]. Mesoscale magnetostructural pattern formation in materials has been explained using Landau expansions of the elastic energy in powers of the strains and the strain gradients [54,56-59], and several key conditions for the formation of mesoscale magnetostructural modulations near structural phase transitions of strongly spin-lattice coupled materials have been delineated [54,60]: (i) a sensitivity of the system to local symmetry-breaking perturbations, *e.g.,* Jahn-Teller instabilities; (ii) the presence of long-range interactions, such as magnetic interactions, that can stabilize particular structural phases locally; and (iii) some local anisotropy, *e.g.,* a surface, defect, or grain boundary, to determine the specific modulation pattern. All of these essential ingredients for the nucleation of mesoscale magnetostructural domain regions are present in both $MnV_2O_4$ and $Mn_3O_4$. It is also worth noting that both $MnV_2O_4$ and $Mn_3O_4$ have orbitally active octahedral (*B*) sites ($V^{3+}$ in $MnV_2O_4$ and $Mn^{3+}$ in $Mn_3O_4$), which has been shown to favor an instability toward spinodal decomposition into coexisting structural phases [61], consistent with our evidence for coexisting tetragonal and orthorhombic phases in $Mn_3O_4$ and similar to earlier evidence for phase coexistence in the Mn-doped spinel $CoFe_2O_4$ [55].

The newest and most significant demonstration from this MFM study is that the mesoscale magnetic domain patterns observed in $MnV_2O_4$ and $Mn_3O_4$ are readily controlled with modest magnetic fields; indeed, the magnetic field strengths at which we observe the magnetic stripe modulations to change in both in $MnV_2O_4$ and $Mn_3O_4$ correspond closely to the magnetic field values at which magnetodielectric effects and magnet-field-tuned lattice striction effects are observed in both $MnV_2O_4$ [30,37] and $Mn_3O_4$ [30,36]. This close correspondence offers strong evidence that the magnetically responsive properties of $MnV_2O_4$ and $Mn_3O_4$ are not associated with homogeneous properties of these materials, but are rather associated with the materials' intrinsic magnetic inhomogeneities, which are ultimately driven by the competition between long-range magnetic interactions and strain energies. Significantly, the presence of domain walls and mesoscale phase separation has been shown to be instrumental in lowering the energy barrier for field-induced phase changes in complex materials [28,29], and indeed, we propose that the mesoscale magnetostructural patterns evident in our MFM results—and their strong susceptibility to magnetic-field manipulation—are primarily responsible for the large magnetic susceptibilities observed in $MnV_2O_4$ [30,37] and $Mn_3O_4$ [30,36].

**Conclusions**

We employed cryogenic MFM and room-temperature EBSD to investigate the nanoscale magnetic properties of the two multiferroic spinel materials $MnV_2O_4$ and $Mn_3O_4$. Our MFM measurements reveal significant nanoscale magnetic domain formation that has been overlooked by previous bulk probe studies. The magnitude of the magnetic modulations in these materials are comparable to the bulk magnetizations measured in these materials, and consequently this nanoscale magnetic inhomogeneity cannot be neglected when considering the overall magnetic behavior of the two materials. The magnetic patterning cannot be attributed solely to simple magnetic domain formation. Theoretical proposals and data interpretations for $MnV_2O_4$ and $Mn_3O_4$ that rely on assumptions of magnetic homogeneity must be revisited. In addition, the presence of nanoscale magnetic inhomogeneity in these two related compounds suggests this phenomenon may be present in other multiferroic spinels.

We have established that mechanical strain plays an important role in the phenomenology of the low-temperature magnetic patterning. In $Mn_3O_4$, the tweed stripe pattern is defined by the tetragonal crystal grains, and stripe pitch is correlated to grain size. In $MnV_2O_4$, the interwoven stripe pattern is also defined by the tetragonal domain structure. When the tetragonal domain structure is determined at experimentally accessible temperatures, we can control the magnetic patterning through application of an external magnetic field. Inducing mechanical strain in $MnV_2O_4$ produces a more complex magnetic pattern at intermediate magnetic fields, and stabilizes magnetic inhomogeneity at higher magnetic fields. These findings are consistent with theoretical results showing that mesoscale magnetic inhomogeneity can significantly lower the energy barrier for strain- and field-dependent phase changes in complex materials, and offers strong evidence that magnetic domain formation plays an important role in the magnetoresponsive behavior of these spinel materials.

## Acknowledgments
Research was supported by the U.S. Department of Energy under Award Number DE-FG02-07ER46453 (B.W., X.W., T.N., and R.B), and by the National Science Foundation under Grants NSF DMR 1464090 and NSF DMR 1800982 (S.L.G. and S.L.C.) and NSF DMR 1455264 (G.J.M). Research at the National High Magnetic Field Laboratory was supported by the National Science Foundation under Grant DMR-1157490, and by the state of Florida. The work was carried out in part in the Frederick Seitz Materials Research Laboratory Central Research Facilities, University of Illinois.

## References
[1] C.-W. Nan, M.I. Bichurin, S. Dong, D. Viehland, and G. Srinivasan, J. Appl. Phys. **103**, 031101 (2008).

[2] H. Katsura, N. Nagaosa, and A.V. Balatsky, "Spin current and magnetoelectric effect in noncollinear magnets," *Physical Review Letters* **95**, 057205 (2005).

[3] D.I. Khomskii, "Multiferroics: different ways to combine magnetism and ferroelectricity," *Journal of Magnetism and Magnetic Materials* **306**, 1 (2006).

[4] S-W. Cheong and M. Mostovoy, "Multiferroics: a magnetic twist for ferroelectricity," *Nature Materials* **6**, 13 (2007).

[5] S. Yang, H.X. Bao, D.Z. Xue, C. Zhou, J.H. Gao, Y. Wang, J.Q. Wang, X.P. Song, Z.B. Sun, X.B. Ren, and K. Otsuka, "Magnetodielectric effect from the onset of ferrimagnetic transition in $CoCr_2O_4$," *Journal of Physics D: Applied Physics* **45**, 265001 (2012).

[6] E.K.H. Salje, "Multiferroic domain boundaries as active memory devices: Trajectories towards domain boundary engineering," *European Journal of Chemical Physics and Physical Chemistry* **11**, 940 (2010).

[7] E.K.H. Salje, O. Aktas, M.A. Carpenter, V.V. Laguta, and J.F. Scott, "Domains within domains and walls within walls: evidence for polar domains in cryogenic $SrTiO_3$," *Physical Review Letters* **111**, 247603 (2013).

[8] G. Lawes, A.P. Ramirez, C.M. Varma, and M.A. Subramanian, "Magnetodielectric effects from spin fluctuations in isostructural ferrimagnetic and antiferromagnetic systems," *Physical Review Letters* **91**, 257208 (2003).

[9] C-H. Yang, T.Y. Koo, and Y.H. Jeong, "How to obtain magnetocapacitance effects at room temperature: the case of Mn-doped $BiFeO_3$," *Solid State Communications* **134**, 299 (2005).

[10] T. Katsufuji and H. Takagi, "Coupling between magnetism and dielectric properties in the quantum paraelectric $EuTiO_3$," *Physical Review B* **64**, 054415 (2001).

[11] U. Adem, G. Nenert, Arramel, N. Mufti, G.R. Blake, and T.T.M. Palstra, "Magnetodielectric coupling by exchange striction in $Y_2Cu_2O_5$," *European Physical Journal B* **71**, 393 (2009).

[12] U. Adem, M. Mostovoy, N. Bellido, A.A. Nugroho, C. Simon, and T.T.M. Palstra, "Scaling behavior of the magnetocapacitance of $YbMnO_3$," *Journal of Physics: Condensed Matter* **21**, 496002 (2009).

[13] N. Mufti, G.R. Blake, T.T.M. Palstra, "Magnetodielectric coupling in $MnCr_2O_4$ spinel," *Journal of Magnetism and Magnetic Materials* **321**, 1767 (2009).

[14] T. Suzuki, K. Adachi, T. Katsufuji, "Coupling between magnetic, dielectric properties and crystal structure in $MnT_2O_4$ (*T*=V,Cr,Mn)," *Journal of Physics: Conferences Series* **31**, 235 (2006).

[15] X. Luo, W.J. Lu, Z.H. Huang, X.B. Hu, L. Hu, X.B. Zhu, Z.R. Yang, W.H. Song, J.M. Dai, and Y.P. Sun, "Large reversible magnetocaloric effect in spinel $MnV_2O_4$ with minimal Al substitution," *Journal of Magnetism and Magnetic Materials* **324**, 766 (2012).

[16] M. Kim, X.M. Chen, Y.I. Joe, E. Fradkin, P. Abbamonte, and S.L. Cooper, "Mapping the magneto-structural quantum phases of $Mn_3O_4$," *Physical Review Letters* **104**, 136402 (2010).

[17] R. Tackett, G. Lawes, B.C. Melot, M. Grossman, E.S. Toberer, and R. Seshadri, "Magnetodielectric coupling in $Mn_3O_4$," *Physical Review B* **76**, 024409 (2007).

[18] T. Suzuki and T. Katsufuji, "Magnetodielectric properties of spin-orbital coupled $Mn_3O_4$," *Physical Review B* **77**, 220402(R) (2008).

[19] G.J. MacDougall, I. Brodsky, A.A. Aczel, V.O. Garlea, G.E. Granroth, A.D. Christianson, T. Hong, H.D. Zhou, and S.E. Nagler, "Magnons and a two-component spin gap in $FeV_2O_4$," *Physical Review B* **89**, 224404 (2014).

[20] S. Lee, H. Takagi, D. Louca, M. Matsuda, S. Ji, H. Ueda, Y. Ueda, T. Katsufuji, J-H Chung, S. Park, S-W Cheong and C. Broholm, "Frustrated magnetism and cooperative phase transitions in spinels," *Journal of the Physical Society of Japan* **79**, 011004 (2010).

[21] G.J. MacDougall et al., "Observation and control of domain wall order in $Mn_3O_4$," unpublished (2017).

[22] K.A. Gschneidner Jr., V.K. Pecharsky and A.O. Tsokol, "Recent developments in magnetocaloric materials," *Reports on the Progress of Physics* **68,** 1479 (2005).

[23] G. Giovannetti, A. Stroppa, S. Picozzi, D. Baldomir, V. Pardo, S. Blanco-Canosa, F. Rivadulla, S. Jodlauk, D. Niermann, J. Rohrkamp, T. Lorenz, S. Streltsov, D. I. Khomskii and J. Hemberger, "Dielectric properties and magnetostriction of the collinear multiferroic spinel $CdV_2O_4$," *Physical Review B* **83**, 060402(R) (2011).

[24] M. Onoda and J. Hasegawa, "A distortion of pseudotetramers coupled with the Jahn-Teller effect in the geometrically frustrated spinel system $CdV_2O_4$," *Journal of Physics: Condensed Matter* **15**, L95 (2003).

[25] P.G. Radaelli, "Orbital ordering in transition metal spinels," *New Journal of Physics* **7**, 53 (2005).

[26] Z.Y. Tian, P.M. Loutou, N. Bahlawane, P.H.T. Ngamou, "Synthesis of the catalytically active $Mn_3O_4$ spinel and its thermal properties," *Journal of Physical Chemistry C* **117**, 6218 (2013).

[27] H. Xia, Y. Wan, F. Yan, and L. Lu, "Manganese oxide thin films prepared by pulsed laser deposition for thin film microbatteries," *Materials Chemistry and Physics* **143**, 720 (2014).

[28] K. H. Ahn, T. F. Seman, T. Lookman, and A. R. Bishop, *Physical Review B* **88**, 144415 (2013).

[29] K. H. Ahn, T. Lookman and A. R. Bishop, *Nature* **428**, 401-404 (2004).

[30] J.-H. Chung, J.-H. Kim, S.-H. Lee, T.J. Sato, T. Suzuki, M. Katsumura, and T. Katsufuji, Phys. Rev. B **77**, 054412 (2008).

[31] G.B. Jensen and O.V. Nielsen, *Journal of Physics C* **7**, 409 (1974).

[32] B. Chardon, and F. Vigneron, *Journal of Magnetism and Magnetic Materials* **58** 128 (1986).

[33] H. D. Zhou, J. Lu, and C. R. Wiebe, "Spin ordering and orbital ordering transitions in $MnV_2O_4$", *Physical Review B* **76**, 174403 (2007).

[34] Y. Nii, H. Sagayama, H. Umetsu, N. Abe, K. Taniguchi, and T. Arima, Phys. Rev. B **87**, 195115 (2013).

[35] V. Hardy, Y. Breard, and C. Martin, Phys. Rev. B **78**, 024406 (2008).

[36] T. Suzuki, M. Katsumura, K. Taniguchi, T. Arima, and T. Katsufuji, Phys. Rev. Lett. **98**, 127203 (2007).

[37] K. Adachi, T. Suzuki, K. Kato, K. Osaka, M. Takata, and T. Katsufuji, *Physical Review Letters* **95**, 197202 (2005).

[38] S.L. Gleason, T. Byrum, Y. Gim, A. Thaler, P. Abbamonte, G.J. MacDougall, L.W. Martin, H.D. Zhou, and S.L. Cooper, *Physical Review B* **89**, 134402 (2014).

[39] K. Takubo, R. Kubota, T. Suzuki, T. Kanzaki, S. Miyahara, N. Furukawa, and T. Katsufuji, Phys. Rev. B **84**, 094406 (2011).

[40] V.O. Garlea, R. Jin, D. Mandrus, B. Roessli, Q. Huang, M. Miller, A.J. Schultz, and S.E. Nagler, Phys. Rev. Lett. **100**, 066404 (2008).

[41] M. Kim, X. M. Chen, X. Wang, C. S. Nelson, R. Budakian, P. Abbamonte, and S. L. Cooper, "Pressure and field tuning the magnetostructural phases of $Mn_3O_4$: Raman scattering and x-ray diffraction studies", *Physical Review B* **84**, 174424 (2011).

[42] Y. Murakami, T. Suzuki, Y. Nii, S. Murai, T. Arima, R. Kainuma, and D. Shindo, "Application of strain to orbital-spin-coupled system $MnV_2O_4$ at cryogenic temperatures within a transmission electron microscope," Microscopy 65, 223 (2016); Y. Murakami, Y. Nii, T. Arima, D. Shindo, K. Yanagisawa, and A. Tonomura, "Magnetic domain structure in the orbital-spin-coupled system $MnV_2O_4$," Physical Review B 84, 054421 (2011).

[43] T. Fukuma, M. Kimura, K. Kobayashi, K. Matsushige, and H. Yamada. "Development of low noise cantilever deflection sensor for multienvironment frequency-modulation atomic force microscopy," *Review of Scientific Instruments* **76**, 053704 (2005).

[44] See Supplemental Material at [URL will be inserted by publisher] for further information on the design of the magnetic probes used for the current experiments and calibration of resultant data.

[45] P.J. Rous, R. Yongsunthon, A. Stanishevsky, and E.D. Williams, "Real-space imaging of current distributions at the submicron scale using magnetic force microscopy: Inversion methodology," *Journal of Applied Physics* **95**, 2477 (2004).

[46] J. Lohau, S. Kirsch, A. Carl, G. Dumpich, and E.F. Wasserman. Quantitative determination of effective dipole and monopole moments of magnetic force microscopy tips," *Journal of Applied Physics* 86 (1999).

[47] T. Goddenhenrich, H. Lemke, M. Muck, U. Hartmann, and C. Heiden, "Probe calibration in magnetic force microscopy," *Applied Physics Letters* **57**, 2612 (1990).

[48] K.L. Babcock, V.B. Elings, J. Shi, D.D. Awshalom, and M. Dugas, "Field-dependence of microscopic probes in magnetic force microscopy," *Applied Physics Letters* **69**, 705 (1996).

[49] L. Kong and S.Y. Chou, "Quantification of magnetic force microscopy using a micronscale current ring," *Applied Physics Letters* **70**, 2043 (1997).

[50] K. Myung-Whun, J.S. Kim, T. Katsufuji, and R.K. Kremer, *Physical Review B* **83** 024403 (2011).

[51] M.C. Kemei, J.K. Harada, R. Seshadri, and M.R. Suchomel, "Structural Change and Phase Coexistence Upon Magnetic Ordering in the Magnetodielectric Spinel $Mn_3O_4$", *Physical Review B* **90**, 064418 (2014).

Wolin, et al. 16

[52] T. Byrum, S.L. Gleason, A. Thaler, G.J. MacDougall, and S.L. Cooper, "Effects of magnetic field and twinned domains on magnetostructural phase mixture in $Mn_3O_4$: Raman scattering studies of untwinned crystals," *Physical Review B* **93**, 184418 (2016).

[53] A. Lavrov, S. Komiya, and Y. Ando, "Magnetic shape-memory effects in a crystal," *Nature* **421**, 230 (2003).

[54] A. Saxena, T. Cast´an, A. Planes, M. Porta, Y. Kishi, T. A. Lograsso, D. Viehland, M. Wuttig, and M. De Graef, "Origin of magnetic and magnetoelastic tweedlike precursor modulations in ferroic materials," *Physical Review Letters* **92**, 197203 (2004).

[55] C.L. Zhang, C.M. Tseng, C.H. Chen, S. Yeo, Y.J. Choi, and S.-W. Cheong, "Magnetic nanocheckerboards with tunable sizes in the Mn-doped $CoFe_2O_4$ spinel," *Applied Physics Letters* **91**, 233110 (2007).

[56] A. M. Bratkovsky, S. C. Marais, V. Heine, and E. K. H. Salje, "The theory of fluctuations and texture embryos in structural phase transitions mediated by strain," *Journal of Physics: Condensed Matter* **6**, 3679 (1994).

[57] A. M. Bratkovsky, E. K. Salje, S. C. Marais, and V. Heine, "Strain coupling as the dominant interaction in structural phase transitions," *Phase Transitions* **55**, 79 (1995).

[58] A. E. Jacobs, "Landau theory of a constrained ferroelastic in two dimensions," *Physical Review B* **52**, 6327 (1995).

[59] A. E. Jacobs, "Landau theory of structures in tetragonal-orthorhombic ferroelastics," *Physical Review B* **61**, 6587 (2000).

[60] X. Wang, "Imaging magnetic order in magnetostructural phases of $Mn_3O_4$," University of Illinois PhD thesis (2012).

[61] M. A. Ivanov, N. K. Tkachev, and A. Y. Fishman, "Phase transformations of the decomposition type in systems with orbital degeneracy," *Low Temperature Physics* **28**, 613 (2002).

# Supplemental Section

## Methods

The magnetic probes used for the experiments detailed in this document are derived from commercially available AFM cantilevers using a unique evaporation process. Cantilevers had varying spring constants and natural frequencies in the ranges $k \sim 0.01\text{-}0.3$ N/m and $f_0 \sim 10\text{-}30$ kHz. Quality factors also varied, centering around $Q \sim 300{,}000$ at $T = 4$K in vacuum. To create magnetic probes suitable for use at high magnetic fields and cryogenic temperatures, we coat only a small portion of the cantilever tip, as showed in Figure S1(a). The cantilever is mounted on a pair of translation stages, which allow the tip to be positioned directly behind a razor blade. A trilayer of Ti-FeCo-Ti is then applied to the exposed portion through electron-beam evaporation. The two Ti layers both promote adhesion and prevent oxidation of the magnetic FeCo layer. We use a 70%-30% mixture of Fe and Co as the magnetic material because it has the maximum saturation magnetization observed in bimetallic alloys [S1]. The coercive field of the MFM probes is typically around $B_C=150$G as measured using a room-temperature testing apparatus.

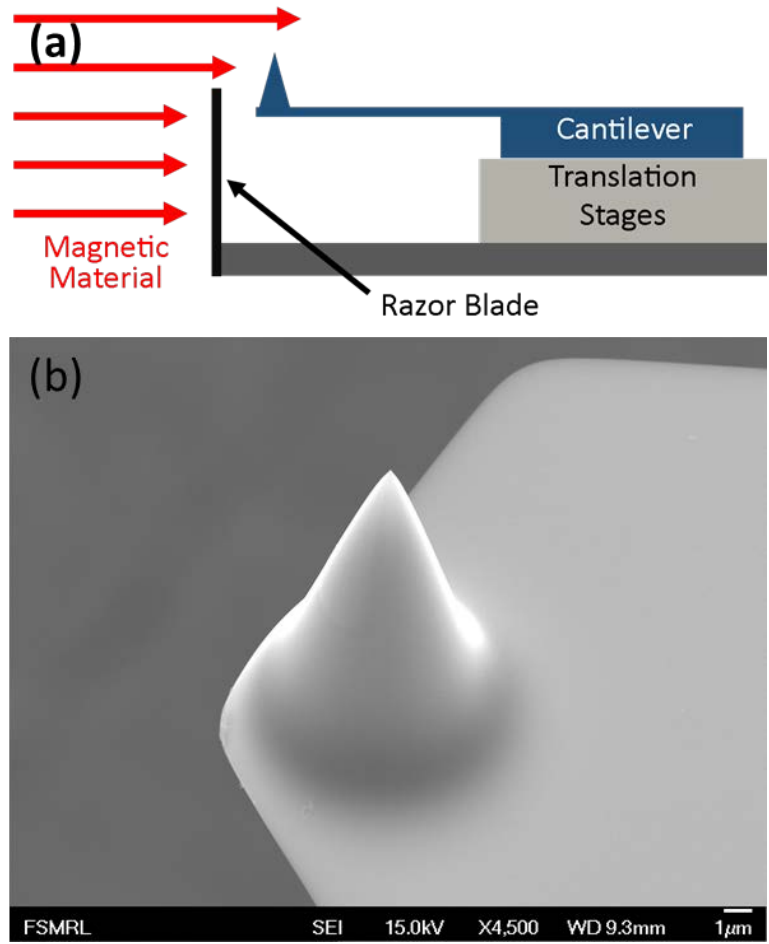

*Figure S1: Magnetic probe preparation. (a) The magnetic probe is created by coating a commercially-available AFM cantilever with magnetic material. Coating only a portion of the cantilever tip prevents unwanted effects caused by high magnetic fields and cryogenic operation. (b) SEM micrograph used as an initial verification that magnetic material has been deposited on the tip. The material appears as an area of light contrast on the upper-left portion of the cone.*

The directional evaporation method we use produces a half-cone thin-film of magnetic material covering approximately half the cantilever height. Figure S1(b) shows an SEM micrograph of the magnetic probe after processing. The trilayer can be seen as an area of light contrast on the upper-left half of the cantilever tip. Our process for creating magnetic probes requires minimal processing and takes good advantage of commercially available products, resulting in a fast turn-around time for magnetic probe production and preservation of the integrity and mechanical properties of the original cantilever.

We used two different cantilevers for the $MnV_2O_4$ measurements: one for the calibrated high-strain experiments and one for the low-strain experiments. The cantilever used for the low-strain experiments had parameters $f = 28.7$ kHz, $Q \approx 200,000$ and $k = 0.23$ N/m at room temperature. The cantilever used for the high-strain experiments had $f = 31.1$ kHz, $Q \approx 125,000$ and $k = 0.31$ N/m at room temperature. Both cantilevers were coated with 10 nm of FeCo for magnetic sensitivity.

We prepared the $MnV_2O_4$ and $Mn_3O_4$ samples using the same polishing regimen. Each rough, as-diced sample was mounted on an aluminum block using Crystalbond epoxy. Using a rotary polishing machine, we manually polished the samples using increasingly fine (1μm, 0.3μm, 0.05μm) alumina polishing powder in a water medium. A final chemi-mechanical polishing step was done using 0.05μm alumina in a basic (pH = 10) medium. For the final step, the sample was placed in a vibratory polishing machine for approximately 6 hours. After polishing the samples were thoroughly cleaned in an ultrasonic immersion cleaner to remove polishing powder and other contaminants. The Crystalbond epoxy was dissolved using acetone, and we mounted the samples as described in the main text.

**Calibration Experiment**

The ability to quantitatively interpret MFM measurements is typically hampered by the lack of information about the nm-scale magnetic details of the magnetic probe [S2,S3]. Without quantitative information about the nm-scale magnetization distribution of the probe, the probe-sample interaction cannot be accurately calculated and modeled. Several effects combine to make *a priori* calculations of the magnetization distribution of the probe difficult, including shape anisotropy effects, thin-film effects, and specifics of the magnetic probe material [S4,S5]. Some techniques exist for measuring the micromagnetic probe structure [S4,S5], but they are time-consuming and still do not provide the necessary accuracy. The preferred option is to use a model magnetic system, such as a current-carrying wire [S6-S9] or calibrated magnetic nanoparticles [S10], to measure the probe response.

The quantity of interest for enabling quantitative MFM analysis is the point spread function (PSF), which describes the probe response to a point-like feature in the magnetic field curvature. By using a magnetic system in which the stray magnetic fields of the sample are known, the point spread function can be extracted by analyzing the known magnetic field distribution and the measured MFM frequency shift data. Following previous work [S2], the frequency response of the MFM probe can be expressed in terms of a 2-dimensional convolution between the magnetic field curvature produced by the sample and the PSF of the tip:

$$\Delta f(\vec{r}_{\parallel}, z) = \frac{f_0}{2k} \int M_z^{(2)}(\vec{r}_{\parallel}' - \vec{r}_{\parallel}) \frac{d^2 B_z}{dz^2}(\vec{r}_{\parallel}', z - h) \, d^2 \vec{r}_{\parallel}'$$

where $f_0$ is the natural frequency of the cantilever, $k$ is the spring constant, $M_z^{(2)}$ is the point spread function, and $h$ is the height above the sample at which the PSF is calculated.

The model magnetic system we used to extract the PSF of our magnetic probe was a 70nm thick current-carrying Au wire lithographically patterned onto a silicon substrate, pictured in Figure S2. The wire had a rectangular cross-section and included various features which have been previously used to calibrate MFM probes [S6-S9]. We found that the junction feature outlined in black (a step-like change in the wire width from 1µm to 4µm) was sufficient to measure the PSF of our magnetic probe. The small marks surrounding the wire are additional lithographically patterned location markers used to locate features on the sample. At either side of the figure, the wire widens where it leaves the measurement region and connects to macroscopic electrodes.

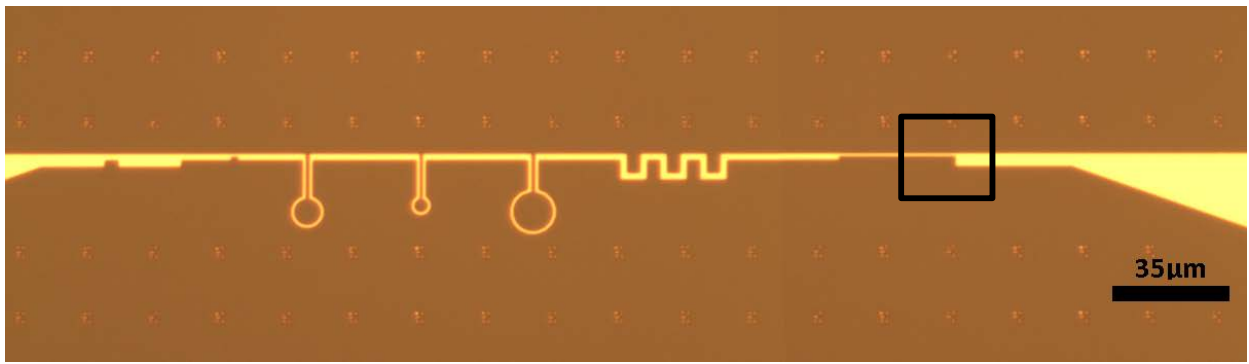

*Figure S2: Composite light micrograph of the MFM calibration sample. The sample is composed of a current-carrying rectangular Au wire. Constrictions, rings, zig-zags, and step junction features were incorporated to have a diverse set of calibration options.*

Figure S3(a) shows MFM frequency shift data for the area surrounding the junction at $T$=4.5K and $B$=20kG. The wire carried $I$=5mA of direct current to produce the known magnetic field distribution. The MFM data show exactly the features we expect for the known magnetic field distribution: opposite-sign frequency shifts centered above the two wire edges and a four-fold increase in the frequency shift magnitudes near the narrower section of the wire. At the inside corner of the junction, there is a peak in the frequency shift, which corresponds to current crowding predicted by the finite element analysis.

We found that the magnetic response of the MFM probe was quantitatively captured by modeling the magnetization distribution as a magnetic point-dipole. This simplified model has been reported as a good approximation for bulk magnetic MFM probes [S6-S9], but has never been investigated for thin-film magnetic probes. Figure S3(b) shows the predicted frequency shift distribution produced by convolving the magnetic field curvature around the junction feature with the point spread function (PSF) of a point-dipole-like magnetic probe (Figure S3(d)). There is excellent qualitative agreement between the simulation results and the MFM data. Two line-cuts across the MFM data are shown in Figure S3(c), along with best-fit results of the point-dipole model obtained by varying the magnetic moment and position relative to the physical tip apex. For this particular MFM probe, the best-fit parameters were $m_z$=2.4·10$^{-15}$ J/T and $h$ = 400nm. Somewhat surprisingly, these values are comparable to those previously reported for bulk magnetic probes, despite the drastic differences in probe materials and geometries [S6,S9]. This similarity further emphasizes the short-range nature of the MFM interaction which limits the effective volume of magnetic material to the region close to the tip apex, where the half-cone thin-film geometry of the probe is less apparent.

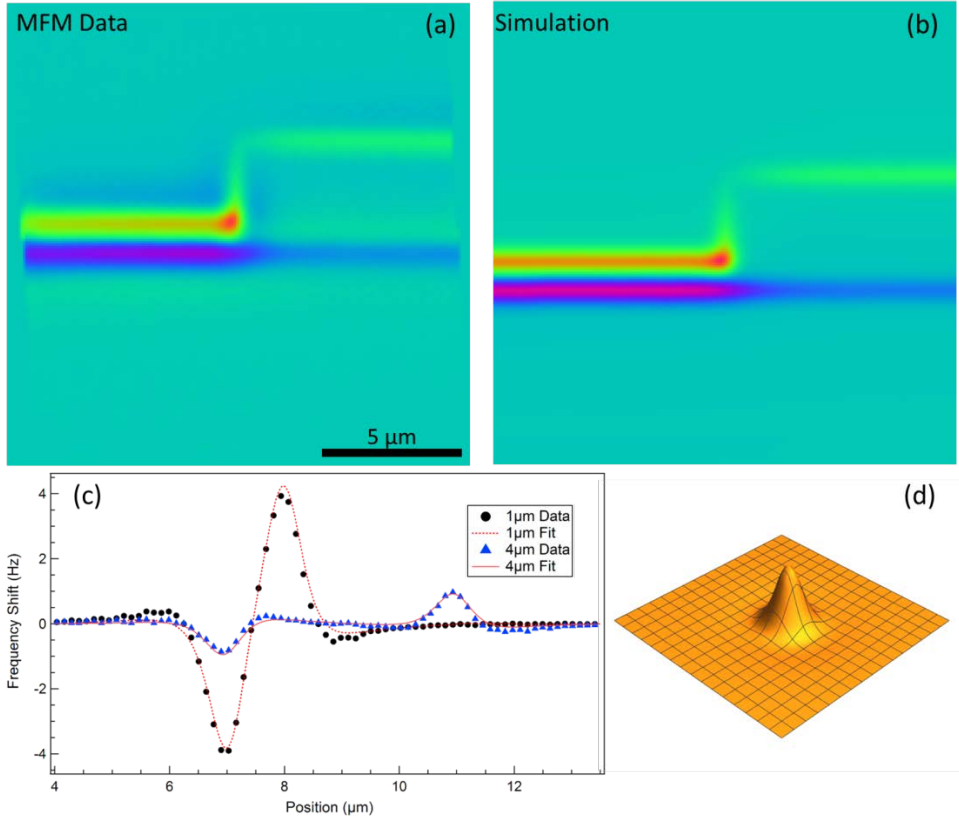

*Figure S3: Calibration experiment results. (a) MFM data collected at T=4K, B=20kG, and I=5mA. (b) The MFM data shows excellent agreement with simulation results for the magnetic field curvature of the wire (obtained through finite element modeling) is convolved with a point-dipole-like point spread function (d). (c) Line cuts through the 1μm and 4μm sections of the wire show quantitative fitting of the model results to the calibration data. The resulting fitting parameter values were used to analyze the MnV$_2$O$_4$ data.*

In order to fully capture the magnetic response of the MFM probe, we recorded data in the presence of various external magnetic fields. The magnetic fields produced by the calibration sample do not vary with external magnetic field, but the magnetic structure and overall magnetization of the MFM probe change with increasing external magnetic field. Between $B$=0kG and $B$=30kG, the effective dipole moment of the magnetic probe increased by approximately 30%, but there was no discernable change in the overall structure of the PSF.

**Modeling of Magnetic Features in MnV$_2$O$_4$**

Using the measured PSF, we can determine the relationship between the magnetic field curvature and the measured frequency shift data, but the ultimate goal is to infer details about the magnetic domain structure of the sample being studied. Following the same strategy we used to fit the point-dipole model to the measured PSF: we generate a proposed magnetic domain structure, calculate the resulting magnetic field curvature, determine the resulting frequency shift pattern by convolving with the PSF, and compare to the experimental data. By changing the parameters of the original proposed domain structure, we can fit the model to the experimental data. Quantitative MFM methods have been previously used to study current crowding effects in artificial systems [S2] and domain characteristics in longitudinal magnetic recording media [S11-S13] but our work is the first application of quantitative MFM to a complex, natural material.

Due to the extended nature of the tip PSF, a single point in the magnetic field curvature produced by the sample interacts with the tip even when the tip is relatively far away (on the order of 1μm). Consequently, sharp changes in the sample magnetization will be smoothed out in the measured frequency data. For the finest stripe features in MnV$_2$O$_4$ (Figure 1(a)), the frequency shift profiles appear approximately sinusoidal, but the underlying sample magnetization is probably much more like a square wave. Square wave and sinusoidal wave magnetization profiles thus provide convenient bounds to our estimates of the sample magnetization. The sharp transitions of a square-wave-like sample magnetization will induce stronger frequency shifts and the smooth transitions of a sinusoidal magnetization will induce weaker frequency shifts for a given sample magnetization.

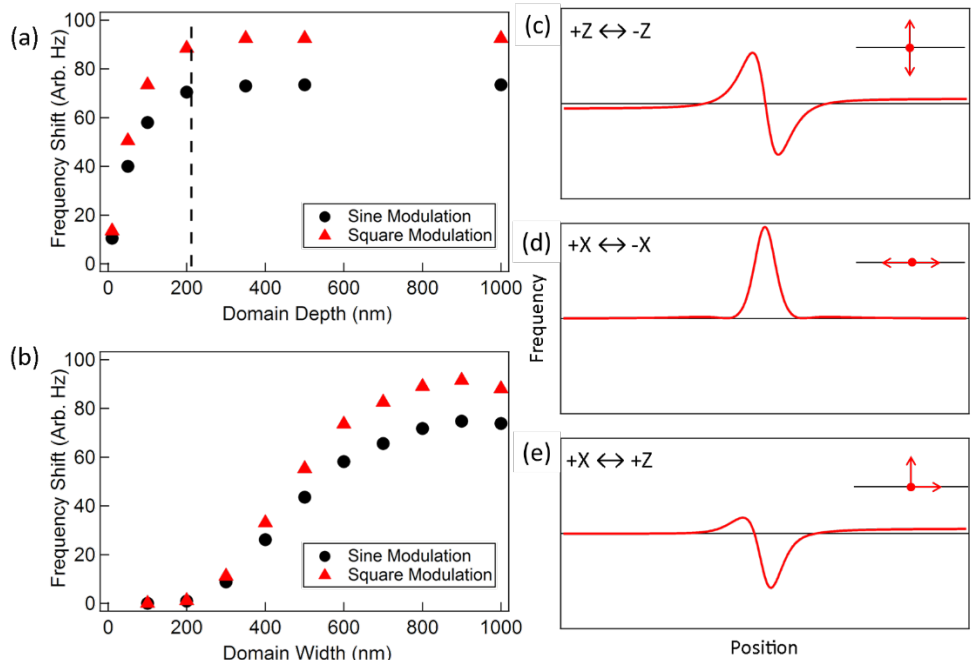

Figure S4: Modeling magnetic domain. (a) The depth of a magnetic domain has a measurable effect on the measured frequency shift only for depths less than approximately the probe-sample separation (dashed line). For domains thicker than this distance, an infinite-thickness approximation is sufficient. (b) Domain width is a crucial parameter in determining the observed frequency shift. For widths less than approximately 900nm, overlap of signals from different features average destructively. (c-e) Different domain wall configurations yield characteristically different frequency shift profiles, which can be used to qualitatively identify domain magnetization direction. The magnetization vectors (red) on either side of the wall as viewed along the sample surface (black) are indicated in the upper right of the respective sub-figure. The three classes of domain wall consist of: (c) 180° change in the magnetization direction across the wall, where both magnetizations are normal to the sample surface; (d) 180° change in the magnetization direction across the wall, where both magnetizations are parallel to the sample surface; and (e) 90° change in the magnetization across the wall, where the magnetization changes from normal to parallel to the sample surface.

Figure S4(a) and (b) show the effects of varying the domain width (in the plane of the sample surface) and depth (into the sample from the surface) on the maximum measured frequency shift for both sinusoidal and square wave stripe profiles. The first important conclusion is that for domains with depth comparable to or greater than the tip-sample separation (in this case $h = 210$nm as indicated by the vertical dashed line in S4(a)) the resulting frequency shift is independent of domain depth. It is highly unlikely that the magnetic domains we observe in $MnV_2O_4$ exist purely at the sample surface, so we conclude that domain depth is not an important parameter in estimating the sample magnetization. However, Figure S4(b) shows a non-trivial variation of the frequency shift according to domain width for the range of stripe feature widths we observe in Figure 1(a), approximately 200-400nm. Note that for features wider than approximately 900nm, the peak frequency decreases slightly. This is the width at which neighboring domains can be resolved by the magnetic probe.

We also used our modeling methods to identify the types of domain walls observed in the low-strain $MnV_2O_4$ sample. Below the Jahn-Teller transition temperature, the tetragonal domains in $MnV_2O_4$ are oriented parallel to one of the three cubic crystal axes, resulting in different types of 90° and 180° domain walls. Furthermore, the MFM probe is sensitive to mainly the z-component of the magnetic field curvature, so both the type of domain wall and the relative orientation of the two magnetizations to the MFM probe determine the resulting frequency shift profile. Figure S4(c) shows the frequency profile of a single 180° domain wall where the two magnetizations point normal to the sample surface. In calculating the estimated domain magnetization in $MnV_2O_4$, we used this type of domain wall. Figure S4(d) shows the frequency profile of a 180° domain wall when the two magnetizations lie within the plane of the sample surface. Due to the anisotropic sensitivity of the probe, this type of domain wall results in a characteristically different frequency profile. Finally, Figure S4(e) shows the frequency profile a 90° domain wall where only one magnetization lies within the plane of sample surface. For certain MFM images of the low-strain $MnV_2O_4$ sample, such as Figure 2(c), we identified the types of domain walls and determined the magnetization orientation in the domains defined by those walls. While beyond the scope of the current work, that analysis can be found elsewhere [S14].

## **Conclusion**

Achieving quantitative magnetic force microscopy results requires an additional step to calibrate the instrument response of the MFM probe. The easiest and most accurate way to calibrate the probe is to measure a known magnetic system and extract the instrument response by comparing the measured frequency shift data to the known magnetic field curvature. For our calibration experiment, we used a current-carrying wire with 2-dimensional features. Additionally, we attempted to exactly reproduce the experimental conditions of our $MnV_2O_4$ measurements during the calibration experiment, including feature sizes, external magnetic fields, and temperatures. We found that the point spread function of the magnetic probe is well-characterized by a point-dipole model. We used further modeling of the magnetic domains $MnV_2O_4$ to account for ambiguities in the magnetization profile and identify the key physical parameters. Finally, we identified the frequency shift profiles associated with different types of domain walls for application to low-strain $MnV_2O_4$ data.

## **References**

[S1] C.W. Chen, *Magnetism and Metallurgy of Soft Magnetic Materials* (North-Holland Publishing, 1977).

[S2] P.J. Rous, R. Yongsunthon, A. Stanishevsky, and E.D. Williams, "Real-space imaging of current distributions at the submicron scale using magnetic force microscopy: Inversion methodology," *Journal of Applied Physics* **95**, 2477 (2004).

[S3] T. R. Albrecht, P. Grutter, D. Horne, and D. Rugar. "Frequency modulation detection using high-Q cantilever for enhanced force microscope sensitivity", *Journal of Applied Physics* **69**, 668 (1991).

[S4] G. Matteucci, M. Muccini, and U. Hartmann, "Electron holography in the study of the leakage field of magnetic force microscope sensor tips," *Applied Physics Letters* **62**, 1839 (1993).

[S5] G. Matteucci, M. Muccini, and U. Hartmann. Flux measurements on ferromagnetic microprobes by electron holography," Physical *Review B* **50**, 6823 (1994).

[S6] J. Lohau, S. Kirsch, A. Carl, G. Dumpich, and E.F.Wasserman. Quantitative determination of effective dipole and monopole moments of magnetic force microscopy tips," *Journal of Applied Physics* 86 (1999).

[S7] T. Goddenhenrich, H. Lemke, M. Muck, U. Hartmann, and C. Heiden, "Probe calibration in magnetic force microscopy," *Applied Physics Letters* **57**, 2612 (1990).

[S8] K.L. Babcock, V.B. Elings, J. Shi, D.D. Awshalom, and M. Dugas, "Field-dependence of microscopic probes in magnetic force microscopy," *Applied Physics Letters* **69**, 705 (1996).

[S9] L. Kong and S.Y. Chou, "Quantification of magnetic force microscopy using a micronscale current ring," *Applied Physics Letters* **70**, 2043 (1997).

[S10] S. Sievers, K.-F. Braun, D. Eberdeck, S. Gustafsson, E. Olsson, H.W. Schumacher, and U. Siegner, "Quantitative measurement of the magnetic moment of individual magnetic nanoparticles by magnetic force microscopy," *Small* **8**, 2675 (2012).

[S11] R. Proksch, G. Skidmore, E.D. Dahlberg, S. Foss, J.J. Schmidt, C. Merton, B. Walsh, and M. Dugas, "Quantitative magnetic field measurements with the magnetic force microscope," *Applied Physics Letters* **69**, 2599 (1996).

[S12] E. T. Yen, H. J. Richter, Ga-Lan Chen and G. Rauch, "Quantitative MFM study on percolation mechanisms of longitudinal magnetic recording," *IEEE Transactions on Magnetics* **33**, 2701-2703 (1997).

[S13] T. K. Taguchi, A. Takeo and Y. Tanaka, "Quantitative MFM study on partial erasure behavior of longitudinal recording," *IEEE Transactions on Magnetics* **34**,1973-1975 (1998).

[S14] B. Wolin, "Real-space magnetic imaging of the spinel $MnV_2O_4$," Ph.D. Dissertation, University of Illinois at Urbana-Champaign (2017).